\documentclass[aps,prd,12pt]{revtex4}
\usepackage[colorlinks=true, pdfstartview=FitV, linkcolor=blue, citecolor=red, urlcolor=magenta]{hyperref}
\usepackage{graphicx}
\usepackage{latexsym}
\usepackage{amsmath}
\usepackage{amsfonts}
\usepackage{amssymb}
\usepackage{verbatim}

\newcommand{\be}{\begin{equation}}
\newcommand{\ee}{\end{equation}}
\newcommand{\bea}{\begin{eqnarray}}
\newcommand{\eea}{\end{eqnarray}}

\newcommand{\pa}{\partial}

\newcommand{\bb}{\bibitem}
\def\pls{\partial\!\!\!/}
\def\bb{\bibitem}
\def\as{A\!\!\!/}

\def\ps{p\!\!\!/}

\def\ds{\partial\!\!\!/}

\def\bb{\bibitem}
\newcommand{\ben}{\begin{eqnarray}}
\newcommand{\een}{\end{eqnarray}}

\begin{document}

\title{Induction of the higher-derivative Chern-Simons extension in QED$_3$ }

\author{$^{1}$M. A. Anacleto}
\email{anacleto@df.ufcg.edu.br}

\author{ $^{1,2}$F. A. Brito}
\email{fabrito@df.ufcg.edu.br}

\author{$^{4}$O.  Holanda}
\email{ozorio.neto@uerj.br}

\author{$^{1,3}$E. Passos}
\email{passos@df.ufcg.edu.br}

\author{$^{2}$A. Yu. Petrov}
\email{petrov@fisica.ufpb.br}

\affiliation{$^{1}$Departamento de F\'{\i}sica, Universidade Federal de Campina Grande,\\
Caixa Postal 10071, 58429-900, Campina Grande, Para\'{\i}ba, Brazil.}
\affiliation{$^{2}$Departamento de F\' isica, Universidade Federal da Para\' iba,\\  Caixa Postal 5008, Jo\~ ao Pessoa, Para\' iba, Brazil.}
\affiliation{$^{3}$Instituto de F\' isica, Universidade Federal do Rio de Janeiro,\\  Caixa Postal 21945, Rio de Janeiro, 
Rio de Janeiro, Brazil.}
\affiliation{$^{4}$Departamento de F\'{i}sica Te\' orica, Instituto de F\' isica, Universidade do Estado do Rio de Janeiro,
Rua S\~ ao Francisco Xavier 524,  20550-013,  Maracan\~ a,  Rio de Janeiro, Brazil}


\begin{abstract}
We perform the perturbative generation of the higher-derivative Chern-Simons contribution to the effective action in the three-dimensional QED at zero and finite temperature. In the latter case, we show that as the temperature goes to infinity this contribution vanishes. However, as expected, as the temperature goes to zero only the covariant part survives.   The non-covariant part contributes only in intermediate temperatures where presents a maximum. 
 
\end{abstract}
 \maketitle


\section{Introduction}
The higher-derivative contributions to the effective action actually call the attention within supersymmetry, gravity, Lorentz-breaking theories and many other contexts. Originally, their study has been motivated by consideration of the gravity models where, because of the non-renormalizability of the Einstein gravity, the hopes for development of the theory consistent at the quantum level have been initially related namely with introduction of the higher derivatives \cite{Stelle}. Despite the difficulties displayed by these theories at the quantum level such as arising of ghosts, see e.g. \cite{Anton},  these theories became an important tool for treating the low-energy effective behaviour, especially in the gravity (for many examples, see the book \cite{BO}), and also in theory of strings and higher spins \cite{strihsp}. Moreover, some efficient manners to deal with these theories have been proposed by Trodden and Fontanini \cite{Trodden} who used a completely Euclidean description for the higher-derivative models, and by Smilga \cite{Smilga} who proposed a way to control the impact of ghosts. However, up to now, most studies for the higher-derivative theories (including supersymmetric and Lorentz-breaking contexts) have been carried out only in four-dimensional space-time whereas their consideration in three dimensions seems to be very interesting -- first, due to the great attention to the three-dimensional field theories inspired by studies of the graphene, second, due to the well-known one-loop finiteness (and, of course, general improvement of the renormalization behavior) and overall simplicity of the three-dimensional field theories. Furthermore alternative higher-derivative gravity {models based on different scalings of space and time coordinates have} been addressed {within} the Hor\v ava-Lifshitz {approach} \cite{Horava:2009uw}. In this case the {higher} derivatives {are present} only in the spatial sector {thus eliminating the ghost problem}, and the theory {turns out to be} consistently {renormalizable} by power counting. {At the same time}, the critical exponent responsible {characterizing the degree of hoigher spatial derivatives} can flow in such a way {that} the theory effectively describes {lower- dimensional physics in high energy limit}. So, {example of Horava-Lifshitz theories} is two-fold. It presents renormalizability of the quantum theory, {and} in the same time {it} effectively flows to a {lower dimensional theory}. Thus, it seems to be important to look for other theories with similar behaviour, i.e. to search for {higher}-derivative theories at lower dimensions. This is the problem we shall address in this paper.

We start with the scheme allowing for inducing a higher derivative Chern-Simons extension via the QED$_{3}$.
The only one such Lorentz-invariant extension is given by \cite{HD,Deser:1988zm}
\bea\label{eq1}
S_{ECS}= (2m )^{-1}\kappa\int d^{3}x \varepsilon^{\alpha\beta\rho}\,  A_{\alpha}\pa_{\beta}\Box A_{\rho}, 
\eea
where $m$ is the mass of theory and $\kappa\propto 1/|m|$ up to a {numerical} factor that we shall determine through radiative induction of this term. Notice that by using integration by parts in the above expression we find
\bea\label{eq2}
S_{ECS}= - (2m )^{-1}\kappa\int d^{3}x \varepsilon^{\alpha\beta\rho}\, \tilde{A}_{\alpha}\pa_{\beta}\tilde{A}_{\rho}\;\;\;{\rm with}\;\;\; \tilde{A}_{\alpha} = \frac{1}{2}\varepsilon_{\alpha\mu\nu}F^{\mu\nu},
\eea
which depends locally on the field strength and {\it not} on the potential.

The main purpose of this paper consists in inducing the term (\ref{eq2}) as a radiative correction through the use of the QED$_{3}$ action by integrating one loop fermionic field at zero and at finite temperature. The paper is organized as follows: {in the section \ref{sec21}, we consider the modified ${\rm QED}_{3}$ and we discuss the gauge invariance and the Ward identities}.  In the section \ref{sec2}, we use the derivative expansion scheme and apply it to induce the higher Chern-Simons actions at zero temperature. {Then we proceed in the sections \ref{sec3}, \ref{sec4} with studying of their} thermal dependence using the Matsubara formalism for fermions. Finally in section \ref{sec5} we present our conclusions.

\section{Classical action of $\rm{QED}_{3}$ with the higher derivative term}\label{sec21}
Let us start with a higher derivative extension to $\rm{QED}_{3}$. Thus, by using the Eq.(\ref{eq2}), we write
\bea\label{eqq1}
\Sigma_{eff}= S_{qed} + S_{ECS} + S_{gf},
\eea
where 
\bea\label{eqq2}
S_{qed}=\int d^{3} x \{\bar{\psi}\big(i\ds - \!m\big)\psi - q\bar\psi \as\psi - \frac{1}{4} F^{\mu\nu}F_{\mu\nu} \}
\eea
is the usual $\rm{QED}_{3}$ contribution (here $\as=\gamma^{\mu} A_{\mu}$, where $A_{\mu}$ is the usual gauge field). The higher derivative term, Eq.(\ref{eq2}), can be rewritten in terms of the field strength as
\bea\label{eqq3}
S_{ECS} = -\frac{\kappa}{4m}\int d^{3}x \,\varepsilon^{\lambda\rho\sigma} F_{\beta\lambda} \pa^{\beta} F_{\rho\sigma}
\eea
and 
\bea\label{eqq4}
S_{gf} = -\frac{1}{2 \xi} \int d^{3}x\, (\pa_{\mu}A^{\mu})^{2}
\eea
is the is a gauge-fixing action (with $\xi$ being the gauge parameter).

To discuss the gauge invariance of the action, we study the corresponding Ward identity. The classical field equations corresponding to Eq.(\ref{eqq1}) are given by
\begin{subequations}
\bea\label{eqq41}
\big(i\overrightarrow{\ds} - \!m \big)\psi(x) = q \as \psi(x),
\eea
\bea\label{eqq42}
\bar\psi(x)\big(- i\overleftarrow{\ds} - \!m \big) = q  \bar\psi(x)\as,
\eea
\bea\label{eqq43}
\Box A_{\mu}(x) - \big(1 - \frac{1}{\xi}\big)\pa_{\mu} \big(\pa_{\nu} A^{\nu}(x)\big) - \kappa(2 m)^{-1}  \varepsilon_{\mu\rho\sigma}\,\Box F^{\rho\sigma}  = q\bar\psi(x) \gamma_{\mu}\psi(x).
\eea
\end{subequations}
Now submitting the effective action (\ref{eqq1}) under the following $U(1)$ gauge transformations
\bea\label{eqq5}
&&\delta A_{\mu}(x) = \frac{1}{q}\pa_{\mu}\omega(x),\nonumber\\
&&\delta\psi(x)=-i\omega(x)\psi(x),\nonumber\\
&& \delta\bar\psi(x)= i\omega(x)\bar\psi(x),
\eea
it transforms according to the Ward identity that writes \cite{bonneau}
\bea
\int d^{3}x \Big\{\frac{1}{q}\pa_{\mu}\omega(x)\frac{\delta \Sigma_{eff}}{\delta A_{\mu}(x)} + i \omega(x) \Big[\bar\psi(x)\frac{\overleftarrow{\delta}\Sigma_{eff}}{\delta\bar\psi(x)} - \frac{\overrightarrow{\delta}\Sigma_{eff}}{\delta\psi(x)}\psi(x) \Big] \Big\}=
- \frac{1}{q \xi}\int d^{3}x\Big\{ \pa_{\mu}A^{\mu} \Box\, \omega(x) \Big\},\nonumber
\eea
so that
\bea
\label{eqq8}
{\it W}_{x}\Sigma_{eff}&\equiv&\pa_{\mu}\frac{\delta \Sigma_{eff}}{\delta A_{\mu}(x)} + iq\Big[\bar\psi(x)\frac{\overleftarrow{\delta}\Sigma_{eff}}{\delta\bar\psi(x)} - \frac{\overrightarrow{\delta}\Sigma_{eff}}{\delta\psi(x)}\psi(x) \Big]\nonumber\\&=&
\frac{1}{\xi} \Box\, \pa_{\mu} A^{\mu}(x).
\eea
Therefore, the presence of the higher derivative term does not modify the Ward identity associated with the usual $\rm{QED}_{3}$.  Notice that the result of the Eq.~(\ref{eqq8}) agrees with the massless regime of the results presented in
\cite{bonneau, Piguet}.

\section{The one-loop effective action}\label{sec2}
In order to perform this calculation we consider the following fermionic part of the QED$_{3}$ Lagrangian 
\bea\label{eq3}
{\cal L}_{\psi}= \bar{\psi}\big(i\ds - \!m\big)\psi - q\bar\psi \as\psi.
\eea
{In principle, our results can be straightforwardly generalized to yield more sophisticated higher-derivative terms, if we replace the $A_{\mu}$ by any vector linear in $A_{\mu}$, such as, for example, $\Box A_{\mu}$, $\partial^{\lambda}F_{\lambda\mu}$, etc., just in this manner some of contributions in \cite{MNP} has been calculated.}
We can integrate out the fermion field $\psi$ in the functional integral to obtain the one loop effective action for the $A_{\mu}$ field that reads
\bea\label{eq04}
S_{eff}[A_{\mu}(x)]= -i{\rm\, Tr\, ln}\big[\ps - m- q \as(x)\big],
\eea
where the symbol $\rm Tr$ stands for the trace over Dirac matrices, trace over the internal space
as well as for the integrations in momentum and coordinate spaces. 
The constant factor has been absorbed into normalization
of the path integral such that
\bea\label{eq05}
S_{eff}[A_{\mu}(x)]=S_{eff}+S^{n}_{eff}[A_{\mu}(x)],
\eea
where
\bea\label{eq06}
S^{n}_{eff}[A_{\mu}(x)]= i{\rm Tr} \sum_{n=1}\frac{1}{n}\Big[ -i q S_{f}(p)\as (x)\Big]^{n}
\eea
and
\bea\label{eq07}
S_{f}(p) = \frac{i}{\ps - m}
\eea
is the usual fermion propagator associated with the theory.

Let us now consider the terms for $n=2$  of the power expanded logarithm in Eq.~(\ref{eq07}) to write
\bea\label{eq08}
S^{n=2}_{eff}[A_{\mu}(x)]=-\frac{i q^{2}}{2}{\rm Tr}\big[S_{f}(p) \as(x) S_{f}(p) \as(x) \big].
\eea
Now applying the main property of derivative expansion method, we observe that any function of momentum can be
converted into a coordinate dependent quantity as \cite{ED}
\bea\label{eq09}
A(x) S_{f}(p) = \big(S_{f}(p -i\pa)A(x)\big).
\eea
The parenthesis on the right hand side merely emphasizes that the derivatives act only on $A(x)$. In this case, the
expression (\ref{eq08}) becomes
\bea\label{eq10}
S^{n=2}_{eff}[A_{\mu}(x)]=-\frac{i q^{2}}{2}{\rm Tr}\big[S_{f}(p) \gamma^{\alpha} \big(S_{f}(p - i \pa)A_{\alpha}\big) \as \big].
\eea
By considering the following expansion
\bea\label{eq11}
S_{f}(p - i \pa) = S_{f}(p) + S_{f}(p) \pls S_{f}(p) +  S_{f}(p) \pls S_{f}(p) \pls S_{f}(p) + S_{f}(p) \pls S_{f}(p) \pls S_{f}(p) \pls S_{f}(p)+ \cdot\cdot\cdot,
\eea
the effective action (\ref{eq10}) can be rewritten in the form
\bea\label{eq12}
S^{n=2}_{eff}[A_{\mu}(x)]=-\frac{i q^{2}}{2} \int d^{3}x \Big[I^{\alpha\mu\nu\lambda\beta}(\pa_{\mu}\pa_{\nu}\pa_{\lambda} A_{\alpha})A_{\beta}\Big],
\eea
where the quantity $I^{\alpha\mu\nu\lambda\beta}$ is given by
\bea\label{eq13}
 I^{\alpha\mu\nu\lambda\beta}=\int \frac{d^{3}p}{(2\pi)^{3}} {\rm tr}[S_{f}(p)\gamma^{\alpha}S_{f}(p)\gamma^{\mu}S_{f}(p)\gamma^{\nu}S_{f}(p)\gamma^{\lambda}S_{f}(p)\gamma^{\beta}]
\eea
and the symbol ${\rm tr}$ denotes the trace of the product of the gamma matrices.
 
Now, we substitute here the expression of the fermion propagator (\ref{eq07})  {together} with the following trace identity: ${\rm tr}[\gamma^{\mu}\gamma^{\nu}\gamma^{\rho}]=2i\varepsilon^{\mu\nu\rho}$. In this case, we can rewrite the expression (\ref{eq13}) as
\bea\label{eq16}
I^{\alpha\mu\nu\lambda\beta}= - 2 m \varepsilon^{\alpha\lambda\beta}\int \frac{d^{3}p}{(2\pi)^{3}}\Big[\frac{\big((p^{2} - m^{2})\eta^{\mu\nu} - 4 p^{\mu}p^{\nu}\big)}{(p^{2} - m^{2})^{4}}\Big].
\eea
The results for relevant momentum integrals, within the dimensional regularization approach, are:
\bea\label{eq19}
\int\frac{d^{3}p}{(2\pi)^{3}}\frac{1}{(p^{2} - m^{2})^{n}}&=& \frac{(-1)^{n}i}{(4\pi)^{3/2}}\frac{\Gamma(n - 3/2)}{\Gamma(n)}\frac{1}{(m^{2})^{n - 3/2}}\nonumber\\&\stackrel{n=3}{\rightarrow}&
\frac{-i}{8\pi}\frac{1}{4m^{2}|m|},
\eea
\bea\label{eq20}
\int\frac{d^{3}p}{(2\pi)^{3}}\frac{p^{\mu}p^{\nu}}{(p^{2} - m^{2})^{n}}&=& \frac{(-1)^{n - 1}i}{(4\pi)^{3/2}}\frac{\eta^{\mu\nu}}{2}\frac{\Gamma(n - 5/2)}{\Gamma(n)}\frac{1}{(m^{2})^{n - 5/2}}\nonumber\\&\stackrel{n=4}{\rightarrow}&
\frac{-i}{8\pi}\frac{\eta^{\mu\nu}}{24m^{2}|m|}.
\eea
As a result, we obtain
\bea\label{eq22}
I^{\alpha\mu\nu\lambda\beta}=\frac{i}{8\pi}\frac{1}{6 m |m|}\varepsilon^{\alpha\lambda\beta}\eta^{\mu\nu}.
\eea
Therefore, the expression (\ref{eq12}) becomes
 \bea\label{eq23}
 S^{n=2}_{eff}[\tilde{A}_{\mu}]= -(2 m)^{-1} \kappa \int d^{3}x \varepsilon^{\alpha\mu\beta}\tilde{A}_{\alpha}\pa_{\mu}\tilde{A}_{\beta},\;\;\;\kappa=\frac{1}{192\pi}\frac{q^{2}}{|m|}. 
 \eea
This is just the term whose form has been proposed in the Introduction.
It is worth to mention that if we abandon the requirement of Lorentz invariance, the structure of the higher-derivative contributions to the effective action can be richer, for example, in the Horava-Lifshitz-like extension of (2+1)-dimensional QED, the higher-derivative contributions to the effective action have been found in \cite{CSHL}, in principle, one can also try to find the three-dimensional analogues of the results obtained in \cite{MNP}, where it is natural to expect the CPT-even Lorentz-breaking terms; however, the action we use essentially differs from that one considered in \cite{CSHL,MNP}.
 
\section{Higher Derivative Chern-Simons at Finite Temperature}\label{sec3}
We shall now make use of imaginary {time} formalism to induce the higher derivative Chern-Simons term at finite temperature. In this case, we change the Minkowski space to Euclidean one by performing the Wick rotation: $x_{0}\to ix_{0}$, $p_{0}\to ip_{0}$ such that $d^{3}x\to id^{3}x$, $d^{3}p\to i d^{3}p$. Thus, we substitute the self-energy tensor (\ref{eq16}) in the effective action (\ref{eq12}) such that the  Euclidean resulted action is given by
\bea\label{eq24}
S^{Euc}_{eff}[A_{\mu}(x)]=i m q^{2} \int d^{3}x \varepsilon^{\alpha\lambda\beta}
\big(\pa_{\mu}\pa_{\nu}\pa_{\lambda}A_{\alpha}\big)A_{\beta}\int \frac{d^{3}p}{(2\pi)^{3}}\Big[\frac{\big((p^{2} + m^{2})\delta^{\mu\nu} - 4 p^{\mu}p^{\nu}\big)}{(p^{2} + m^{2})^{4}}\Big].
\eea
To develop calculations with finite temperature, let us now assume that the system is
in the state of the thermal equilibrium with a temperature $T=1/\beta$. In this case we can
use the Matsubara formalism for fermions. This consists of taking $p_{0}\to \omega_{n}=(n + 1/2)2\pi/\beta$ and replacing the integration over zeroth component of the momentum by a discrete sum {over $n$}: $(1/2\pi)\int dp_{0}\to 1/\beta\sum\limits_{n}$ \cite{mat}. Additionally, we implement translation only on the space coordinates of the loop momentum $p_{\mu}$ and we
decompose it as follows: $p_{\mu}\to \vec{p}_{\mu} + p_{0}\delta_{0\mu}$ \cite{lvtf}. This {leads} us to the identity
\bea\label{eq25}
\vec{p}_{\mu}\vec{p}_{\nu}\to\frac{\vec{p}^{2}}{2}\big(\delta_{\mu\nu}-\delta_{0\mu}\delta_{0\nu}\big).
\eea
Here, we use the covariance under spatial rotations. Now, we apply all the above information in the effective action (\ref{eq24}), therefore {the temperature-dependent effective action turns out to}
reproduce the following structure:
\bea\label{eq26}
S^{Euc}_{eff}[A_{\mu}(x)]&=&\frac{i m q^{2}}{\beta} \int d^{3}x \Bigg[\varepsilon^{\alpha\lambda\beta}\big(\pa^{2}\pa_{\lambda}A_{\alpha}\big)A_{\beta} \sum_{n=-\infty}^{+\infty}\int \frac{d^{2}\vec{p}}{(2\pi)^{2}}\Big[\frac{1}{(\vec{p}^{2} + M_{n}^{2})^{3}} - \frac{2\vec{p}^{2}}{(\vec{p}^{2} + M_{n}^{2})^{4}}\Big]-\nonumber\\&&4\varepsilon^{\alpha\lambda\beta}\big(\pa_{0}^{2}\pa_{\lambda}A_{\alpha}\big)A_{\beta} \sum_{n=-\infty}^{+\infty}\int \frac{d^{2}\vec{p}}{(2\pi)^{2}}\Big[\frac{\frac{\vec{p}^{2}}{2} - M_{n}^{2} + m^{2}}{(\vec{p}^{2} + M_{n}^{2})^{4}}\Big]\Bigg].
\eea 
where 
\bea\label{eq222}
M_{n}^{2}= \big(n + \frac{1}{2}\big)^{2} \frac{4 \pi^{2}}{\beta^{2}} + m^{2}.
\eea
After performing the momentum integration, we {arrive at}
\bea\label{eq27}
S^{Euc}_{eff}[A_{\mu}(x)]&=&\frac{i q^{2}}{16\pi^{2}}\frac{a^{3}}{m |m|}\int d^{3}x\Bigg[ \frac{1}{3}\varepsilon^{\alpha\lambda\beta}\big(\pa^{2}\pa_{\lambda}A_{\alpha}\big)A_{\beta} \sum_{n=-\infty}^{+\infty}\frac{1}{\big[(n + \frac{1}{2})^{2} + a^{2}\big]^{2}} +\nonumber\\&+& 2\varepsilon^{\alpha\lambda\beta}\big(\pa_{0}^{2}\pa_{\lambda}A_{\alpha}\big)A_{\beta}\times\nonumber\\&&\Bigg[\sum_{n=-\infty}^{+\infty}\frac{1}{\big[(n + \frac{1}{2})^{2} + a^{2}\big]^{2}} - \frac{4}{3}\sum_{n=-\infty}^{+\infty}\frac{a^{2}}{\big[(n + \frac{1}{2})^{2} + a^{2}\big]^{3}}\Bigg]\Bigg],
\eea
where $a=\frac{m\beta}{2\pi}$. Now, we use the {well-known} results for {sums} 
over the Matsubara frequencies {(see f.e. \cite{Ford})}:
\bea\label{eq28}
&&\sum_{n=-\infty}^{+\infty}\frac{1}{\big[(n + \frac{1}{2})^{2}+a^{2}\big]^{2}}=
-\frac{\pi}{a^{3}}\bigg[\pi a\;{\rm sech}^{2}(\pi a) - {\rm tanh}(\pi a)\bigg],\nonumber\\&&\sum_{n=-\infty}^{+\infty}\frac{a^{2}}{\big[(n + \frac{1}{2})^{2}+a^{2}\big]^{3}}=
\frac{\pi}{8a^{3}}\bigg[3 {\rm tanh}(\pi a)- \pi a\;{\rm sech}^{2}(\pi a)\big(3 + 2\pi a\;{\rm tanh}(\pi a)\big)\bigg]
\eea
into the expression (\ref{eq27}) to find
\bea\label{eq29}
S^{Euc}_{eff}[A_{\mu}(x)]= \frac{i q^{2}}{8\pi}\frac{1}{6 m |m|}\int d^{3}x \Big[
\varepsilon^{\alpha\lambda\beta}\big(\pa^{2}\pa_{\lambda}A_{\alpha}\big)A_{\beta}\, f(a) +\varepsilon^{\alpha\lambda\beta}\big(\pa_{0}^{2}\pa_{\lambda}A_{\alpha}\big)A_{\beta}\, g(a)\Big],
\eea
where the quantities $f(a)$ and $g(a)$ are `thermal functions' given respectively by
\bea\label{eq30}
&&f(a)= -\pi a\;{\rm sech}^{2}(\pi a) + {\rm tanh}(\pi a),\nonumber\\&&
g(a)= 2(\pi a)^{2} {\rm tanh}(\pi a)\,{\rm sech}^{2}(\pi a).
\eea 
\begin{figure}[!h]
	\centering
		\includegraphics[width=7.0cm,height=6.5cm]{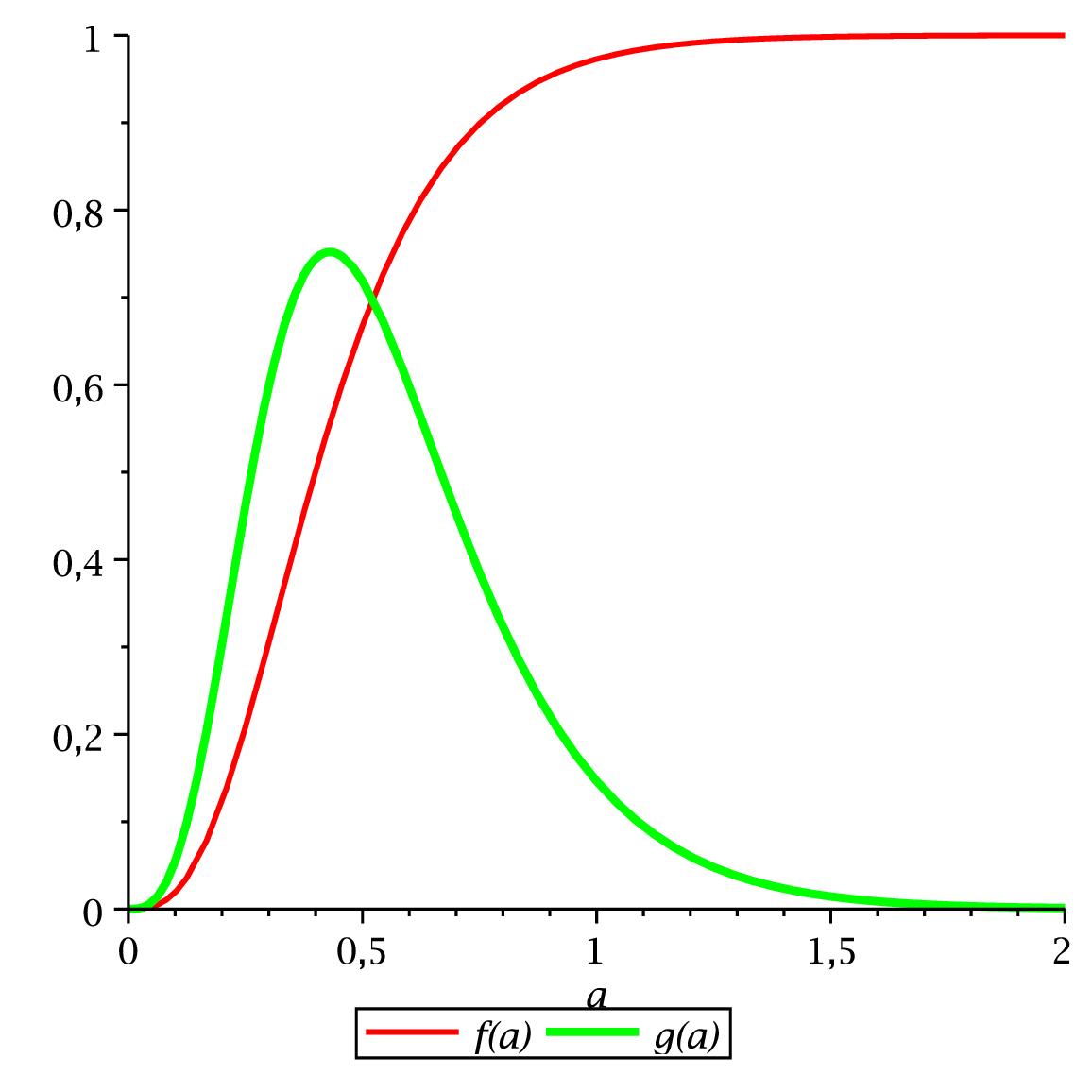}
	\caption{The behavior of $f(a)$ and $g(a)$, corresponding to covariant and non-covariant contributions, respectively. Since $a=\frac{m\beta}{2\pi}$, as the temperature is very high these functions vanish. As expected, at very low temperature only the covariant part survives. The non-covariant contribution presents a maximum in intermediate temperatures.}
	\label{fig1}
\end{figure}
They comprise the covariant and non-covariant contributions, respectively. The asymptotic analysis of the above thermal functions {yields} the following results
\bea\label{eq31}
&& f(a\to 0)\to 0\;\;(T\to \infty),\;\;\;f(a\to \infty)\to 1\;\;(T\to 0);\nonumber\\
&& g(a\to 0)\to 0\;\;(T\to \infty),\;\;\;f(a\to \infty)\to 0\;\;(T\to 0).
\eea
The full behavior is depicted in Fig.~\ref{fig1}. Recalling that $a=\frac{m\beta}{2\pi}$, we see that as the temperature is very high these functions vanish. However, as the temperature goes to zero only the covariant part of (\ref{eq29}) survives and recovers the result obtained {for} zero temperature (\ref{eq23}), as expected. {It is interesting to note that} the non-covariant part contributes only in intermediate temperatures presenting a maximum {for some intermediate temperature}.
\section{Thermal Effective Action}\label{sec4}
The resulted (\ref{eq28}) {displays} the possibility to construct a thermal higher-derivative theory. The {associated quantum correction can now be written as the gauge fixing term, so, in the Minkowski space the effective Lagrangian is}
\bea\label{eq32}
{\cal L}_{thm}= -\frac{1}{4}F_{\mu\nu}F^{\mu\nu} + \frac{\tilde\kappa}{2m} \varepsilon^{\mu\beta\nu}(\pa_{\beta}A_{\mu})\big(\pa^{2}f(a) + \pa_{0}^{2} g(a)\big)A_{\nu}- \frac{1}{2\xi} (\pa\cdot A)^{2},
\eea
where $\tilde\kappa= q^{2}/24 \pi |m|$. Here, {we want to use}  the Lagrangian (\ref{eq32}) to investigate the effects of temperature on the gauge excitations. A way of study this matter is {based on the} associated propagator. 
Thus, we can {immediately write the form of our propagator in usual form corresponding to the Maxwell-Chern-Simons action (in Feynman gauge $\xi=1$)},
\bea\label{eq33}
\Delta^{\mu\nu}= -\big[p^{2} - m^{2}\big]^{-1} \big(g^{\mu\nu} - i (m/p^{2}) \varepsilon^{\mu\alpha\nu}p_{\alpha}\big),
\eea
{with the following redefinition $m \to \tilde\kappa/m \big(p^{2} f(a) + p_{0}^{2} g(a)\big) $ is carried out} in the momentum space. The resultant {thermal propagator is given} by
\bea\label{eq34}
\Delta^{\mu\nu}_{th}&=&\Big[ p^{2} - (m^{2})^{-1}\big(p^{2} f(a) + p_{0}^{2} g(a)\big)^{2} {\tilde\kappa}^{2}\Big]^{-1}\Big(g^{\mu\nu} - i (m p^{2})^{-1} \times\nonumber\\&\times&
(p^{2} f(a) + p_{0}^{2} g(a)\big)\tilde\kappa\, \varepsilon^{\mu\alpha\nu}p_{\alpha}\Big).
\eea
Notice that the denominator  associated to propagator (\ref{eq34}) describes two groups of gauge excitations,
\bea\label{eq35}
\frac{1}{\Big[ p^{2} - (m^{2})^{-1}\big(p^{2} f(a) + p_{0}^{2} g(a)\big)^{2} {\tilde\kappa}^{2}\Big]}= -\frac{1}{p^{2}} + \frac{1}{p^{2}\big(1 - \frac{m^{2} p^{2}}{\big(p^{2} f(a) + p_{0}^{2} g(a)\big)^{2} {\tilde\kappa}^{2}}\big) }
\eea
which one group is massless and the other are massive {and thermal} given by
\begin{eqnarray}\label{36}
E_{\pm}= \pm\frac{\sqrt{\lambda_{1}\pm \sqrt{\lambda_{1}^{2} - \lambda_{2} (h(a))^{2}}}}{h(a)}
\end{eqnarray}
where $h(a)=f(a)+g(a)$ and the quantities: $\lambda_{1}$ and $\lambda_{2}$ are given by
\begin{eqnarray}\label{37}
\lambda_{1}&=& f(a)h(a) {\mid\vec{p}\mid}^2 + \frac{\tilde{m}^2}{2}\\
\lambda_{2}&=& {\mid\vec{p}\mid}^2 \big({\mid \vec{p}\mid}^2 + \tilde{m}^2\big),\;\;\;{\rm with}\;\;\;\tilde{m}=m/\kappa.
\end{eqnarray}
Notice that in the limit of temperature goes to zero $(T\to 0)$ the massive-gauge excitations are restored.
However, in the massless case, the gauge excitations remain dependent on thermal functions. 

\section{Conclusion}\label{sec5}

We have showed how the higher-derivative terms can be generated at the one-loop order in the three-dimensional QED at zero and finite temperature. In the latter case, we have considered a covariant and non-covariant part. They {display different behaviour}. Whereas the covariant part is recovered as the temperature goes to zero, the non-covariant part has a maximum only in intermediate temperatures. On the other hand, both parts go to zero as the temperature goes to infinity. The fact that the covariance is {strongly} broken by the non-covariant part in some intermediate scale of high temperature (see Fig.~\ref{fig1}) may set a scale of covariance breaking in theories where spatial and temporal momentum components scales differently, such as in Ho\v rava-Lifshitz gravity \cite{Horava:2009uw}. 

The natural continuation of the study we performed here could consist in generalization of our results for non-Abelian and Lorentz-breaking cases (it worth to mention that, up to now, the number of known Lorentz-breaking results in three-dimensional space-time is very small). Also, this methodology can be applied as well to generating of higher-derivative corrections in linearized gravity and to obtaining more sophisticated quantum corrections in the Horava-Lifshitz-like case generalizing the results of \cite{CSHL}.


{\acknowledgments} We would like to thank to CNPq, PNPD-CAPES, PROCAD-NF/2009-CAPES for partial financial support. The work by A. Yu. P. has been partially supported by the CNPQ project 303783/2015-0.


\end{document}